\documentclass[
reprint,
superscriptaddress,
amsmath,
amssymb,
aps]{revtex4-2}

\usepackage{xcolor}
\definecolor{mygreen}{HTML}{0a8154}
\definecolor{myblue}{RGB}{0,91,210}
\definecolor{mybluee}{HTML}{1987b1}
\definecolor{myred}{HTML}{bf0000}
\definecolor{mymaroon}{HTML}{800000}

\usepackage[
colorlinks=true,
linkcolor=red,
citecolor=myblue,
urlcolor=mymaroon,
linkcolor=myred]{hyperref}

\usepackage{graphicx}
\usepackage{orcidlink}
\usepackage{xcolor}
\usepackage{placeins}

\usepackage{tikz}
\usetikzlibrary{arrows,calc,positioning}
\tikzstyle{intt}=[draw,text centered,minimum size=6em,text width=5.25cm,text height=0.34cm]
\tikzstyle{intl}=[draw,text centered,minimum size=2em,text width=2.75cm,text height=0.34cm]
\tikzstyle{int}=[draw,minimum size=2.5em,text centered,text width=3.5cm]
\tikzstyle{inti}=[draw,minimum size=3em,text centered,text width=5cm]
\tikzstyle{intii}=[draw,minimum size=3em,text centered,text width=8cm]
\tikzstyle{intg}=[draw,minimum size=3em,text centered,text width=7cm]
\tikzstyle{ints}=[draw,minimum size=3em,text centered,text width=2.5cm]
\tikzstyle{sum}=[draw,shape=circle,inner sep=2pt,text centered,node distance=3.5cm]
\tikzstyle{summ}=[drawshape=circle,inner sep=4pt,text centered,node distance=3.cm]

\begin{document}
	
	\title{Constraining Twin-Star Existence in Four-Parameter Maxwell Equation of State}

	\author{Shamim Haque \orcidlink{0000-0001-9335-5713}}
	\email[Contributed equally]{}
	\affiliation{Indian Institute of Science Education and Research Bhopal, Bhopal 462066, India}
	\affiliation{Indian Institute of Astrophysics, Block 2, 100 Feet Road, Koramangala, Bengaluru 560034, India}
	
	\author{Atharva Shinde \orcidlink{0009-0006-8431-3072}}
	\email[Contributed equally]{}
	\affiliation{Indian Institute of Science Education and Research Bhopal, Bhopal 462066, India}
	
	\author{Asim Kumar Saha \orcidlink{0009-0000-8375-4833}}
	\affiliation{Indian Institute of Science Education and Research Bhopal, Bhopal 462066, India}
	
	\author{Tuhin Malik \orcidlink{0000-0003-2633-5821}}
	\affiliation{Departamento de Física, Universidade de Coimbra, 3004-516 Coimbra, Portugal}
	
	\author{Ritam Mallick \orcidlink{0000-0003-2943-6388}}
	\email{mallick@iiserb.ac.in}
	\affiliation{Indian Institute of Science Education and Research Bhopal, Bhopal 462066, India}
	
	\date{\today}
	\begin{abstract}
		Twin stars (TSs) are predicted to exist in nature if the hadron-to-quark phase transition is strong enough to form a new branch of hybrid stars, separated from the branch of neutron stars. It is essential to study the hybrid equation of states (EoSs) that can form TSs, with a primary focus on estimating the onset of the phase transition and its strength. We systematically study the TS scenario by adopting a four-parameter Maxwell-CSS approach to construct a large ensemble of Maxwell-type hybrid EoSs. Within this framework, we report the conditions on our parameter space imposed by the TOV solutions. For a fixed transition energy density and speed of sound for quark matter, we define distinct ranges of transition pressures based on the allowed strengths of discontinuity. Below a maximum transition pressure, a range of discontinuity exists that increases as the transition pressure decreases. Thereby, we identify the loci of the limits on discontinuities as the TS existence boundaries. Based on the causality limit, the TS existence boundaries can be truncated. Strong constraints on this picture are drawn from the CET, GW170817 and NICER. Within our considered model, we computed conditional extrema surviving the adopted $2\sigma$ filters---the maximum mass for TSs to be $2.05~M_\odot$, the allowed strongest discontinuity in rest-mass density to be $7.76\rho_\mathrm{sat}$, and the upper bound on transition rest-mass density to be $4.03\rho_\mathrm{sat}$. Subsequently, we compute the implications of the stiffness of the quark matter EoS on this picture. Different confidence levels for observational constraints are considered to assess the extent to which Maxwell-type EOSs remain compatible with the observations.
	\end{abstract}
	
	\maketitle
	
	\section{Introduction}
	\label{sec:intro}
	
	The pursuit of understanding the equation of state (EoS) of matter inside a Neutron Star (NS) is a traditional one~\cite{kumar_theoretical_2024,lattimer_neutron_2021,chatziioannou_neutron_2025}. It remains an important revisit as we continue to observe more of the universe with increasingly advanced technologies, particularly by measuring these compact objects and witnessing the violent events involving them. NSs possess strong gravity such that the matter at the core is compressed beyond nuclear saturation density $\rho_\mathrm{sat} \simeq 2.7\times10^{14}~\mathrm{g/cm}^{3}$ \cite{glendenning_compact_1997}. The unavailability of such circumstances in Earth-based environments makes NSs the natural laboratories in the universe to study the behaviour of matter at low temperatures and high densities \cite{lattimer_physics_2004,annila_neutron_2023,burgio_neutron_2021,baym_hadrons_2018}. At such extreme conditions, a phase transition (PT) from hadronic matter (HM) to quark matter (QM) is expected from Quantum Chromodynamics (QCD) \cite{shuryak_quantum_1980}. Violent astrophysical events, such as supernova explosions \cite{sagert_signals_2009,zha_impact_2022,khosravi_largani_constraining_2024,jakobus_role_2022,kuroda_core-collapse_2022,fischer_core-collapse_2011,fischer_quark_2018,zha_progenitor_2021,huang_phase-transition-induced_2025} and binary NS mergers \cite{Most:2018eaw,Weih:2019xvw,prakash_detectability_2024,espino_revealing_2024,blacker_constraining_2020,bauswein_identifying_2019,ujevic_reverse_2023,kedia_binary_2022,huang_merger_2022,fujimoto_gravitational_2023,fujimoto_signature_2025,haque_effects_2024,oechslin_influence_2004,chatterjee_distinct_2026,Hensh:2024onv}, are predicted to reveal such phenomena.   
	
	The objective of achieving an integrated picture of QCD has been approached from various directions. In the principles of Chiral Effective Field Theory (CET) \cite{keller_nuclear_2023,tews_constraining_2018,drischler_how_2020,drischler_chiral_2019,hebeler_equation_2013}, the EoS of the cold matter is well-anticipated to be in hadronic state upto $\sim 1.1\rho_\mathrm{sat}$ \cite{alford_relativistic_2022,drischler_chiral_2021,lonardoni_nuclear_2020,holt_equation_2017,hebeler_nuclear_2015,drischler_microscopic_2014,machleidt_chiral_2011,epelbaum_modern_2009}. From the earth-based experiments involving the relativistic heavy-ion collision has mainly probed the EoS of matter at high temperatures and low densities \cite{adams_experimental_2005,arsene_quarkgluon_2005,back_phobos_2005,adcox_formation_2005,gehrmann_precision_2022}. Theoretical framework of lattice QCD \cite{bazavov_equation_2009,borsanyi_full_2014,nagata_finite-density_2022} efficiently probes the matter at low densities, however, interfered by sign problem \cite{troyer_computational_2005,goy_sign_2017} when extended to densities at the regimes of NSs. Perturbative QCD (pQCD) calculations \cite{andersen_gluon_2010, andersen_qcd_2011,gorda_cool_2022,mogliacci_equation_2013,gorda_equation_2023,gorda_soft_2021,gorda_next--next--next--leading_2018,kurkela_cool_2016,kurkela_cold_2010,haque_three-loop_2014,haque_next--next-leading-order_2021,ghiglieri_perturbative_2020} have computed the behaviour of matter at asymptotically high densities ($\sim 40\rho_\mathrm{sat}$).

	To probe the density regime that resides inside the core of an NS ($1\sim10~\rho_\mathrm{sat}$), there are a couple of directions to move forward. Strong constraints have been drawn from the precise mass-radius measurements made by the NICER X-ray observations  \cite{riley_nicer_2019,miller_psr_2019,raaijmakers_nicer_2019,riley_nicer_2021,miller_radius_2021,raaijmakers_constraints_2021,fonseca_refined_2021,mauviard_nicer_2025,choudhury_nicer_2024}. Importantly, the recent mass-radius measurement of PSR J0614--3329 \cite{mauviard_nicer_2025} and PSR J0437--4715 \cite{choudhury_nicer_2024} point towards a softer EoS, in contrast to the earlier measurements of J0030+0451 \cite{riley_nicer_2019,miller_psr_2019,raaijmakers_nicer_2019} and J0740+6620 \cite{riley_nicer_2021,miller_radius_2021,raaijmakers_constraints_2021,fonseca_refined_2021}, which suggested otherwise. Additional assistance came from the multi-messenger astronomy, mainly the observation of gravitational waves from binary merger of NSs \cite{abbott_gw170817_2017}, inferring constraints on the tidal deformability of NSs \cite{raithel_tidal_2018,de_tidal_2018,rezzolla_using_2018,bauswein_neutron-star_2017,annala_gravitational-wave_2018,abbott_gw170817_2018}, which also steers the investigation to a softer composition of matter around the canonical mass of NS ($1.4~M_\odot$). On the other hand, the mass measurement of massive pulsars ($\gtrsim2.0~M_\odot$) \cite{ozel_massive_2010,cromartie_relativistic_2020,antoniadis_massive_2013,arzoumanian_nanograv_2018,demorest_two-solar-mass_2010} advocate for a relatively stiffer EoS.
	
	Despite the progress made, there can still exist a plethora of thermodynamically consistent EoSs that comply with the observational status discussed above \cite{gorda_constraints_2023,gorda_ab-initio_2023,somasundaram_investigating_2023,lindblom_spectral_2012,annala_evidence_2020}. The attempts to probe the properties of PT remain important yet inadequate. Primarily, the onset of PT is not known a priori, and the nature of first-order PT is proposed by models that include---($i$) Maxwell construction--the PT is modelled as a transition using a discontinuity jump in the density at a constant pressure, ($ii$) Gibbs construction--the HM and QM coexist in a mixed phase inside a region of pressure-density \cite{glendenning_phase_2001,glendenning_first-order_1992}. With such prospects and uncertainties in hand, model-agnostic approaches are useful for encapsulating all possibilities in an EoS describing the behaviour of matter inside NSs \cite{altiparmak_sound_2022,annala_evidence_2020,read_constraints_2009,oboyle_parametrized_2020,chatterjee_analyzing_2024}.
	
	A typical hybrid EoS with Maxwell-type PT constructed out of an agnostic approach consists of a pure hadronic part at lower densities and a pure quark part at higher densities, separated by a PT region described by a discontinuous jump in the density at constant transition pressure. Under the general relativistic hydrostatic equilibrium conditions governed by the Tolman-Oppenheimer-Volkoff (TOV) equations \cite{tolman_static_1939,oppenheimer_massive_1939}, such EoSs indicate the existence of hybrid stars (HSs), which contain a quark seed within a hadronic shell. Interestingly, within the constraints imposed by astrophysical observations, the discontinuity in the density jump that describes the strength of the PT can be large enough to give rise to a third branch of HSs, separated from the second branch of NSs \cite{alford_generic_2013,christian_classifications_2018}. In this context, the first branch of compact objects refers to the family of white dwarfs. Such scenarios form the twin stars (TSs) that reside on the third branch (smaller radius, more compact) but have equal mass to NSs on the second branch (larger radius, less compact).

	TSs are the direct manifestation of strong hadron-to-quark PT. Observation of such candidates of compact objects can confirm the occurrence of PT in the astrophysical scenarios and severely constrain the EoS \cite{zacchi_twin_2017,gorda_constraints_2023,mendes_constraining_2025,montana_constraining_2019,tsaloukidis_twin_2023,christian_twin_2020,christian_confirming_2022,li_confronting_2025,landry_prospects_2022,alvarez-castillo_properties_2025,grundler_bayesian_2025,li_hybrid_2024,christian_which_2024,blomqvist_strong_2026,alvarez-castillo_third_2019}, especially if detections are made in the high-mass regime \cite{benic_new_2015,goncalves_electrically_2022,christian_supermassive_2021,alvarez-castillo_high-mass_2017}. On the other hand, the formation of TSs in realistic scenarios and their dynamical evolution remain a serious pursuit \cite{espino_fate_2022,naseri_exploring_2024,chan_distinct_2026}. Several studies have presented the possibility of realistic EoSs that can construct such species \cite{alford_compact_2017,zhang_impact_2025,jakobus_possibility_2021,zacchi_twin_2017,sharifi_studying_2021,sen_detailed_2022,laskos-patkos_signatures_2023,zhou_dwarf_2023,zhou_hidden_2026,yang_hybrid_2026,laskos-patkos_hybrid_2024,li_ultracompact_2023,zhang_self-bound_2026,ayriyan_bayesian_2026}. A few works have studied the oscillation modes \cite{pradhan_probing_2024,goncalves_fundamental-mode_2022,rau_two_2023,rodriguez_classification_2026,zhu_probing_2023,laskos-patkos_binding_2026,pereira_phase_2018}, modified gravity \cite{lope-oter_twin_2025,lope-oter_constraining_2024}, QCD trace anomaly \cite{jimenez_how_2024}, rotation \cite{bhattacharyya_rotating_2005,dimmelmeier_dynamic_2009,bejger_consequences_2017,negreiros_rotational_2025}, magnetic fields \cite{gomes_can_2019,sotani_quark_2017}, temperature \cite{lyra_compactness_2023,carlomagno_thermal_2024,carlomagno_hybrid_2024} and dark matter effects \cite{kumar_stability_2025,das_influence_2026} on these compact objects. TSs have also been studied for slow stable HSs \cite{ranea-sandoval_constraining_2022, lugones_model-agnostic_2023, ranea-sandoval_asteroseismology_2023, mariani_could_2024, mariani_how_2025, mariani_systematic_2026}.
	
	En route to uncovering the TS scenario, it is essential to sketch an extensive (if not complete) picture of TSs, considering the large possibility of hybrid EoSs that can coexist with recent observational constraints. We adopt a four-parameter Maxwell-CSS approach to construct a large ensemble of Maxwell-type hybrid EoSs and systematically obtain TS solutions within this framework. In particular, transition energy density and transition pressure are independently varied to allow a larger set of hadronic EoSs within the causality limit to form hybrid EoSs. Within this framework, a systematic analysis of EoSs reveals more of the TS scenario than the previous literature. After establishing the framework under hydrostatic equilibrium within the limits of causality, we constrain the picture using recent observational inferences through hard cut filtering and set bounds on the properties of PT and the maximum mass of TSs. In particular, the recent observations of PSR J0614--3329 and PSR J0437--4715 are rather strong and constitute an important constraint to be included in new studies.
	
	Importantly, we revisit the criteria for TS solutions as given by the TOV equations to gain an extensive picture beyond the existing literature. In particular, we define the TS existence boundaries for a choice of transition energy density and stiffness of the quark matter EoS. The effects of causality and the properties of PT on these curves are exhaustively addressed, revealing specific conditions that can give rise to truncated TS existence boundaries. Previous studies have constructed TS existence boundaries only for a limited or fixed hadronic EoS and thus have missed this incompleteness of the TS existence boundary arising from the causality constraint \cite{montana_constraining_2019,christian_classifications_2018,alford_generic_2013}. 
	
	The article is organised in the following way. In Sec.~\ref{sec:form}, we discuss the four-parameter Maxwell-CSS framework adopted to construct the ensemble of Maxwell-type hybrid EoS, define our parameter space, and outline our strategy for locating the third branch. In Sec.~\ref{sec:Results}, we discuss our results in three main parts---approach from TOV, constraints from observations, and possible bounds on the TS scenario. Finally, we conclude our work with remarks in Sec.~\ref{sec:Summ}. Since this article is a collection of many nested cases based on various EoS parameters being varied or kept fixed, we have given a roadmap of the flow of the discussions along with indications to the respective subsections and associated figures/tables.
	
	\begin{figure*}
		\centering
		\begin{tikzpicture}
			\node [inti] (kp0)  {{\underline{4-Parameter EoS Construction}}\vspace*{1mm}\\$P_\mathrm{tr}, ~e_\mathrm{tr},~\Delta e,~c^2_\mathrm{s,QM}$\\{\it Section \ref{sec:eos}, Figure \ref{fig:formalism_1}}};
			
			\node [inti] (kp1) [node distance=0.5cm,right=of kp0] {{\underline{Inferences from TOV}}\vspace*{1mm}\\Emergence of TS existence boundaries\\Varying $P_\mathrm{tr}$\\Fixed $e_\mathrm{tr},~c^2_\mathrm{s,QM}$\\{\it Section \ref{sec:theory}, Figure \ref{fig:theory_1}}};
			
			\node [inti] (kp2) [node distance=0.5cm,right=of kp1] {{\underline{Effect on TS Existence Boundary}}\vspace*{1mm}\\Varying $e_\mathrm{tr},~c^2_\mathrm{s,QM}$\\{\it Section \ref{sec:theory}, Figure \ref{fig:theory_2}}};
			
			\node [intii] (kp3) [node distance=0.5cm,below=of kp1] {{\underline{Observational Constraints on TS Existence Boundary}}\vspace*{1mm}\\Varying $e_\mathrm{tr}$ ($2\rho_\mathrm{sat}$, $3\rho_\mathrm{sat}$, $4\rho_\mathrm{sat}$)\\Fixed $c^2_\mathrm{s,QM}$=$1.0$\\{\it Section \ref{sec:obs}, Table \ref{tab:parameter_ranges}}, Figures \ref{fig:observation_1} and \ref{fig:observation_2}};
			
			\node [intg] (kp30) [node distance=0.7cm and -3.4cm,below left=of kp3] {
				Varying observational conf.~($2\sigma$ and $3\sigma$)\\{\it Section \ref{sec:obs}}\\{\it Figures \ref{fig:observation_3} (top to bottom) and \ref{fig:observation_4}}};
			
			\node [intg] (kp31) [node distance=0.7cm and -3.4cm,below right=of kp3] {
				Varying $c^2_\mathrm{s,QM}$ (0.55, 0.62, 0.7, 0.85, 1.0)\\{\it Section \ref{sec:obs}}\\{\it Figures \ref{fig:observation_3} (left to right) and \ref{fig:observation_4}}};
			
			\node [ints] (kp300) [node distance=0.8cm and -1.5cm,below left=of kp30] {
				Finding $e^\mathrm{max}_\mathrm{tr}$\\Fixed $c^2_\mathrm{s,QM}$=1.0\\{\it Section \ref{sec:limits}}\\{\it Table \ref{tab:e_tr_max}}};
			
			\node [ints] (kp301) [node distance=0.8cm and -4.4cm,below left=of kp30] {
				Finding $\Delta e_\mathrm{max}$\\Fixed $c^2_\mathrm{s,QM}$=1.0\\{\it Section \ref{sec:limits}}\\{\it Table \ref{tab:del_e}}};
			
			\node [ints] (kp302) [node distance=0.8cm and -7.3cm,below left=of kp30] {
				Finding $M^\mathrm{max}_\mathrm{TS}$\\Fixed $c^2_\mathrm{s,QM}$=1.0\\{\it Section \ref{sec:limits}}\\{\it Table \ref{tab:m_max}}};
			
			\node [ints] (kp310) [node distance=0.8cm and -7.3cm,below right=of kp31] {
				Finding $e^\mathrm{max}_\mathrm{tr}$\\Fixed $2\sigma$ conf.\\{\it Section \ref{sec:limits}}\\{\it Table \ref{tab:e_tr_max_cs2}}};
			
			\node [ints] (kp311) [node distance=0.8cm and -4.4cm,below right=of kp31] {
				Finding $\Delta e_\mathrm{max}$\\Fixed $2\sigma$ conf.\\{\it Section \ref{sec:limits}}\\{\it Table \ref{tab:del_e_max_cs2}}};
			
			\node [ints] (kp312) [node distance=0.8cm and -1.5cm,below right=of kp31] {
				Finding $M^\mathrm{max}_\mathrm{TS}$\\Fixed $2\sigma$ conf.\\{\it Section \ref{sec:limits}}\\{\it Table \ref{tab:m_max_cs2}}};
			
			\draw[->] (kp0) -- (kp1);
			
			\draw[->] (kp1) -- (kp2);
			
			\draw[->] (kp2) |- (kp3);
			
			\draw[->] (kp3) -- (kp30);
			\draw[->] (kp3) -- (kp31);
			
			\draw[->] (kp30) -- (kp300);
			\draw[->] (kp30) -- (kp301);
			\draw[->] (kp30) -- (kp302);
			
			\draw[->] (kp31) -- (kp310);
			\draw[->] (kp31) -- (kp311);
			\draw[->] (kp31) -- (kp312);
			
		\end{tikzpicture}
		\caption{Roadmap of the flow of discussions and nested cases in the article, along with indications to the respective sections and associated figures/tables.}
		\label{fig:flowchart}
	\end{figure*}

	\section{Formalism}
	\label{sec:form}
	
	\subsection{Equation of State}
	\label{sec:eos}
	
	We employ a four-parameter Maxwell-CSS model to systematically construct EoSs that provide a macroscopic description of matter properties, particularly the connection between the thermodynamic quantities---rest-mass density ($\rho$), energy density ($e$), and pressure ($P$). The EoSs are computed using the following parameters:
	\begin{itemize}
		\item {\it transition energy density} ($e_\mathrm{tr}$)
		\item {\it transition pressure} ($P_\mathrm{tr}$)
		\item {\it discontinuity in energy density} ($\Delta e$)
		\item {\it constant speed of sound for quark EoS} ($c^2_\mathrm{s,QM}$)
	\end{itemize}
	The independent choices of $e_\mathrm{tr}$ and $P_\mathrm{tr}$ decide the stiffness of hadronic EoS. These parameters are sufficient to describe Maxwell-type EoSs, which mainly involve large discontinuities. We recall that TSs have been categorised based on the maximum mass of the second branch ($M_\mathrm{TOV,2}$) and the third branch ($M_\mathrm{TOV,3}$) in Ref.~\cite{christian_classifications_2018}, which are as follows:
	\begin{itemize}
		\item {\it Category I}: $M_\mathrm{TOV,2}\geq2.0~M_\odot$ and $M_\mathrm{TOV,3}\geq2.0~M_\odot$, particularly associated with higher $e_\mathrm{tr}$. 
		\item {\it Category II}: $M_\mathrm{TOV,2}\geq2.0~M_\odot$ and $M_\mathrm{TOV,3}<2.0~M_\odot$, particularly associated with larger $\Delta e$.
		\item {\it Category III}: $1.0~M_\odot \leq M_\mathrm{TOV,2}<2.0~M_\odot$ and $M_\mathrm{TOV,3}\geq2.0~M_\odot$.
		\item {\it Category IV}: $M_\mathrm{TOV,2}<1.0~M_\odot$ and $M_\mathrm{TOV,3}\geq2.0~M_\odot$, particularly associated with low $P_\mathrm{tr}$.
	\end{itemize}
	See Fig.~3 in Ref.~\cite{montana_constraining_2019} for schematic reference. Each EoS is set up using the four sections explained below:
	\begin{itemize}
		\item {\it Crust Region}: At smaller densities ($\lesssim 0.4\rho_\mathrm{sat}$), a fixed crust EoS is built using four piecewise polytropes, which are taken from Table II in Ref.~\cite{read_constraints_2009}. The addition of this part remains common throughout all the EoSs constructed for this study.
		\item {\it Hadronic Region} ($e<e_\mathrm{tr}$): An additional polytrope is added above the crust EoS to mimic the hadronic matter, with adiabatic index $\gamma$ as a free parameter. The thermodynamic quantities---$P$ and $e$ are given in terms of $\rho$ as,
		\begin{equation}\label{eq:had}
			P(\rho)= k\rho^\gamma,\quad e(\rho)= \rho (a+1)+ \dfrac{P}{\gamma-1}
		\end{equation}

		\begin{figure*}
			\centering
			\includegraphics[scale=1]{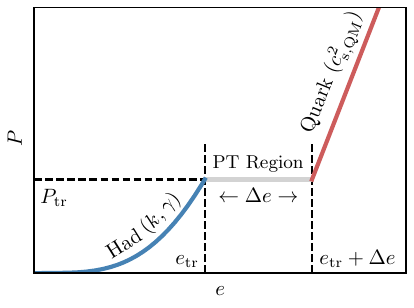}\hspace{0.5cm}
			\includegraphics[scale=1]{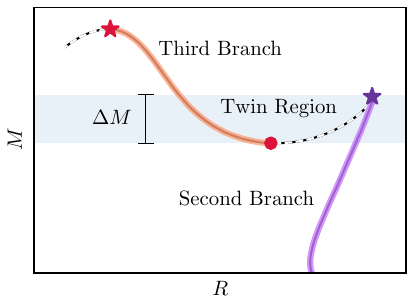}
			
			\caption{[Left] Diagram indicating the different parts of the EoS that are controlled by different parameters---transition energy density ($e_\mathrm{tr}$), transition pressure ($P_\mathrm{tr}$), the strength of the discontinuity in energy density ($\Delta e$) and the constant speed of sound for the quark matter EoS ($c^2_\mathrm{s,QM}$). [Right] Diagram of the $M$--$R$ sequence of an EoS that constructs TSs. The second (purple) and third (orange) stable branches are separated by a sequence of unstable branches (black dashed curve). The maximum mass on the second and third branches ($M_\mathrm{TOV,2}$ and $M_\mathrm{TOV,3}$) is indicated by purple and red stars, respectively. The sequence of unstable stars begins beyond these stellar configurations. The minimum mass on the third branch is indicated by a red circle. The twin region is defined as the mass range in which equal-mass stellar models are present on both stable branches.}
			\label{fig:formalism_1}
		\end{figure*}
		
		The polytropic constant ($k$) and $a$ are computed using the boundary condition to ensure continuity of $P(\rho)$ and $e(\rho)$ between the crust and the hadronic EoS. The independent choices for $e_\mathrm{tr}$ and $P_\mathrm{tr}$ allow for finding an appropriate value of $\gamma$ and transition rest-mass density ($\rho_\mathrm{tr}$). These quantities are numerically computed using the equations,
		\begin{equation}
			\rho_\mathrm{tr}=\left(\dfrac{P_\mathrm{tr}}{k}\right)^{{1}/{\gamma}} , \quad e_\mathrm{tr}= \rho_\mathrm{tr}(1+a) + \dfrac{P_\mathrm{tr}}{\gamma-1}
		\end{equation}
		For a fixed $e_\mathrm{tr}$, a higher (lower) value of $P_\mathrm{tr}$ will set a higher (lower) value of $\gamma$, resulting in a stiffer (softer) hadronic EoS.
		\item {\it Phase Transition Region} ($e_\mathrm{tr}<e<e_\mathrm{tr}+\Delta e$): The PT region is computed after fixing $e_\mathrm{tr}$ and $P_\mathrm{tr}$, allowing a free choice to be made for $\Delta e$. This value decides the strength of the discontinuity of PT. The pressure will remain constant in this region as given by $P_\mathrm{tr}$. This step leads to the construction of Maxwell-type PT \cite{glendenning_first-order_1992}. Since the chemical potential remains constant in the region of PT, the $\Delta\rho$ can be computed as,
		\begin{equation}
			\Delta\rho = \dfrac{\rho_\mathrm{tr}\Delta e}{P_\mathrm{tr}+e_\mathrm{tr}}
		\end{equation}

		\item {\it Quark Region} ($e>e_\mathrm{tr}+\Delta e$): The quark part is modelled using the constant speed of sound (CSS) parametrisation with $c_\mathrm{s,QM}^2$ \cite{alford_generic_2013,christian_classifications_2018}. The pressure in the quark region is given as,
		\begin{equation}
			P(e)= P_\mathrm{tr} + c_\mathrm{s,QM}^2 (e - e_\mathrm{tr} - \Delta e)
		\end{equation}
	\end{itemize}
	
	Figure~\ref{fig:formalism_1} (left) shows the schematic diagram of the EoS construction method described above. First, we fix a value of $e_\mathrm{tr}$, after which the $P_\mathrm{tr}$ can be varied from an upper bound set by the limit of causality; that is, the maximum value of $P_\mathrm{tr}$ is set at a value where $c^2_\mathrm{s}(P_\mathrm{tr},e_\mathrm{tr})=1.0$. However, if the central pressure of $M_\mathrm{TOV}$ of hadronic EoS is smaller than this upper limit of pressure, then the maximum value of $P_\mathrm{tr}$ is set to the central pressure of the stellar configuration at $M_\mathrm{TOV}$. This rule assures that we induce PT in the relevant positions of an $M$--$R$ sequence and not in the cores of unstable stellar models beyond $M_\mathrm{TOV}$.

	On the other hand, the lower limit of $P_\mathrm{tr}$ is set to a value where such a central pressure would construct an NS of $1.0~M_\odot$, which means $M_\mathrm{TOV,2}> 1.0~M_\odot$. However, if an unstable branch is formed by an EoS that constructs  $M_\mathrm{TOV,2}=1.0~M_\odot$, we only increase $\Delta e$ up to a limit where the third branch forms a minimum mass up to $0.7~M_\odot$. We do not construct EoSs which form a minimum mass below $0.7~M_\odot$ in the third branch. These choices are made to restrict our search to reasonable PT densities. In theory, such EoS (with large $\Delta e$ and very small $P_\mathrm{tr}$) can be constructed where TSs can form at very low masses ($\lesssim0.5~M_\odot$), which are far from the NS range (minimum mass of NSs is known to be $\sim 1.1~M_\odot$~\cite{martinez_pulsarj04531559_2015,muller_minimum_2025}). Searches for such TS solutions are avoided in this study. We do not survey the TSs of category IV, which are not responsible for constructing massive TSs or assisting in setting upper limits on the $e_\mathrm{tr}$ and $\Delta e$. Moreover, small $P_\mathrm{tr}$ at large $e_\mathrm{tr}$ will make extremely soft hadronic EoSs, resulting in hybrid EoSs that will be unable to support massive NSs ($\gtrsim 2.0~M_\odot$). Hence, we find TSs of category IV will be incompatible at large $e_\mathrm{tr}$, within the considered model. At lower $e_\mathrm{tr}$, the acceptance of such TSs can be anticipated from our analysis and does not require specific exploration. See Refs.~\cite{christian_classifications_2018,zhang_impact_2025,jakobus_possibility_2021,lope-oter_twin_2025,laskos-patkos_hybrid_2024}, which have especially examined TSs of category IV.
	
	We construct a bunch of hadronic EoSs by uniformly varying $P_\mathrm{tr}$ in the described range and set the quark matter EoS on the other side by fixing a value for $c^2_\mathrm{s,QM}$.  The $\Delta e$ (and hence  $\Delta \rho$) is adaptively varied to find the limits on its allowed range, such that the given pair of hadronic and quark EoS can construct TS solutions. The accuracy of determining the limits on $\Delta e$ ($\Delta \rho$) is considered up to $10^{12}~\mathrm{g/cm}^3$ ($\approx 0.01\rho_\mathrm{sat}$).

	\begin{figure*}
		\centering
		\includegraphics[scale = 0.95]{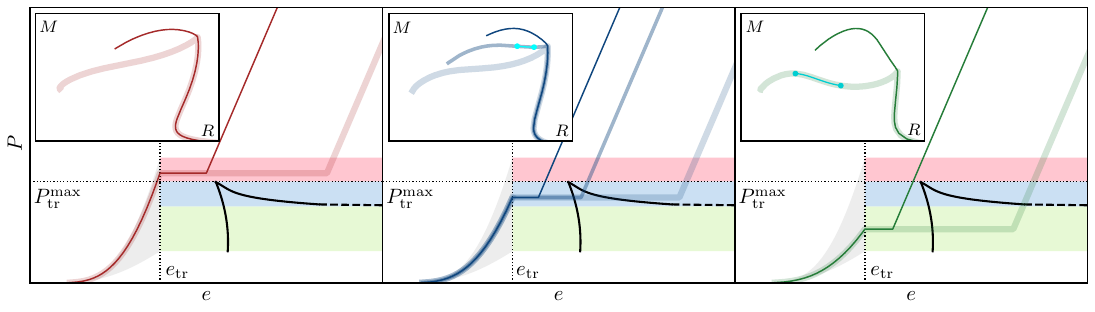}
		\caption{Allowed region of EoS space with fixed $e_\mathrm{tr}$ that forms TS solutions. The grey region indicates the range of hadronic EoS considered in constructing the Maxwell-type hybrid EoSs. The {\it solution-less region} (red) indicates the range of $P_\mathrm{tr}$ inside which TSs do not form. The left panel shows the $M$--$R$ sequences of two EoSs in this region with different $\Delta e$. The {\it bound region} (blue) is separated from the red region using $P^\mathrm{max}_\mathrm{tr}$, at which a $\Delta e$ exists that allows TS solutions. The middle panels show that the EoS with $\Delta e$ within the allowed values of $\Delta e$ constructs TSs. The sequence of the TSs is indicated by cyan in the inset. The {\it unbound region} (green) only contains the minimum value of $\Delta e$ above which TS solutions are possible. The right panel shows that the EoS with a large value of $\Delta e$ forms TSs. The sequence of TSs is indicated by turquoise in the inset. The loci of minimum and maximum $\Delta e$ are indicated in black, which we identify as the TS existence boundary. The black dashed line is an extrapolation of the maximum $\Delta e$.}
		\label{fig:theory_1}
	\end{figure*}

	\subsection{Locating third branch}
	\label{sec:ts}
	
	Figure~\ref{fig:formalism_1} (right) shows the schematic diagram of the $M$--$R$ sequence that has separated the second and third branches containing TSs. To detect the TS solutions in the third branch, we first locate the central density of the maximum mass configuration on the second branch (purple curve), which is constructed from the hadronic EoS. Based on the Seidov criteria \cite{seidov_stability_1971}, upon increasing the central density to build further stellar models, the EoS may allow for two scenarios. The immediate sequence of HSs may continue to be in the second branch, or they may fall into the unstable branch (discussed as scenarios ($b$) and ($d$) in Fig.~2 of reference \cite{alford_generic_2013}). In either of the situations, the $M_\mathrm{TOV,2}$ is computed as indicated with a purple star in Fig.~\ref{fig:formalism_1} (right). Consequently, an unstable branch (black dashed curve) is recognised only if the mass of the subsequent stellar models decreases up to $\Delta M =0.001~M_\odot$ with respect to the purple star. The quark EoS can be stiff enough that a third branch (orange curve) may appear after a sequence of unstable HSs. The third branch is detected only if the mass of the subsequent stellar models increases up to $\Delta M = 0.001~ M_\odot$ from the minimum mass in the unstable branch (indicated with a red circle). Finally, the third branch reaches its $M_\mathrm{TOV,3}$ (indicated with a red star), and the subsequent stellar models form another unstable branch  (black dashed curve). Once we numerically obtain this sequence of branches within the tolerance $\Delta M$, the EoS is recognised to have produced TS solutions. 
	
	The detection margin of $\Delta M=0.001~M_\odot$ is suitable for our study, as this accuracy is several orders of magnitude above the observational accuracy \cite{lattimer_neutron_2019,lattimer_nuclear_2012,huxford_accuracy_2024,mauviard_nicer_2025,choudhury_nicer_2024,riley_nicer_2019,miller_radius_2021,fonseca_refined_2021,antoniadis_massive_2013,ozel_massive_2010,cromartie_relativistic_2020,arzoumanian_nanograv_2018,demorest_two-solar-mass_2010,doroshenko_strangely_2022,kini_constraining_2024}. Besides, detection margins $ \Delta M < 0.001~M_\odot$ lead to ambiguity between the physical stable/unstable branches versus a local fluctuation due to numerical calculation, which are highly probable when TOV equations have to handle EoSs with strong and discontinuous PT.

	\section{Results} 
	\label{sec:Results}
	
	We discuss the results from two main perspectives. Within the described parameter space, we analyse the conditions on allowed EoSs that give rise to the TS scenario, purely governed by TOV solutions. Based on the established theoretical framework, we apply recent observations to constrain the transition densities, the strength of the PT discontinuity, and the maximum mass of TSs.
	
	\subsection{Inferences from TOV solutions}
	\label{sec:theory}
	
	Figure~\ref{fig:theory_1} shows a schematic picture of the region in EoS space with a fixed $e_\mathrm{tr}$, indicating the allowed region of $\Delta e$ that can form TSs. The grey region indicates the range of hadronic EoSs constructed according to the fixed $e_\mathrm{tr}$ and the allowed range of $P_\mathrm{tr}$ set by our formalism. By varying $P_\mathrm{tr}$ in this range and fixing a value for $c^2_\mathrm{s,QM}$, we identify three distinct regions based on the allowed conditions on $\Delta e$, which are explained below.

	\begin{figure*}
		\centering
		\includegraphics[scale = 1]{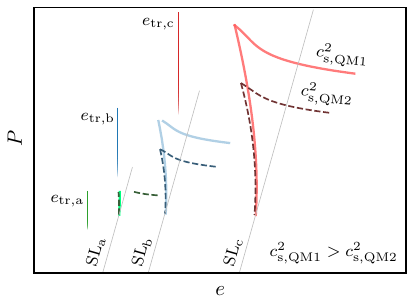}\hspace{0.5cm}
		\includegraphics[scale = 1]{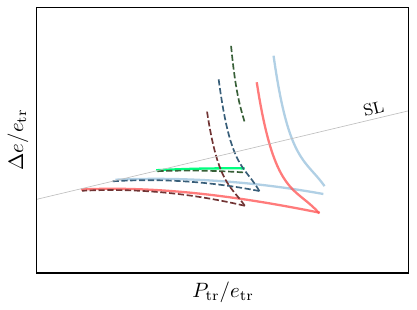}
		\caption{The effects of change in $e_\mathrm{tr}$ and $c^2_\mathrm{s,QM}$ on TS existence boundary. [Left] The vertical fading lines indicate the $e_\mathrm{tr}$. The respective Seidov lines of each $e_\mathrm{tr}$ are indicated in grey solid lines. [Right] The Seidov lines align in the modified axes.}
		\label{fig:theory_2}
	\end{figure*}

	{\it Solution-less Region (Red)}---In this region, TSs do not form for any value of $\Delta e$, as shown in Fig.~\ref{fig:theory_1} (left panel). The inset shows the $M$--$R$ sequences of selected EoSs that have $P_\mathrm{tr}$ in this region, emphasising that a third branch does not appear in these sequences after entering the unstable branch beyond $M_\mathrm{TOV,2}$. For such combinations of $e_\mathrm{tr}$ and $P_\mathrm{tr}$, the quark EoS is not stiff enough to form another stable branch beyond the unstable branch.
	
	{\it Bound Region (Blue)}---The maximum value of $P_\mathrm{tr}$ in this region (indicated as $P^\mathrm{max}_\mathrm{tr}$) is extracted numerically, at which TSs can form for a unique value of $\Delta e$. For $P_\mathrm{tr}< P^\mathrm{max}_\mathrm{tr}$, there exists a range of $\Delta e$, in which TSs can form, as shown in Fig.~\ref{fig:theory_1} (middle panel). We selected three EoSs in this region, two of which lie outside the allowed region of $\Delta e$. The inset showing the respective $M$--$R$ sequences indicates that only the EoS having $\Delta e$ within the allowed region forms a third branch beyond the second branch, which is separated by an unstable branch. The sequence of TSs is indicated by the cyan curve on the third branch. Decreasing $P_\mathrm{tr}$ results in an increase in the maximum limit of $\Delta e$ and thereby increases the range of $\Delta e$ allowed for TS solutions. At the lowest value of $P_\mathrm{tr}$ in this region, the maximum limit of $\Delta e$ is no longer found in our numerical setup. However, an extrapolation using the locus of maximum $\Delta e$ in this region indicates the growing (rather than blowing up) nature of maximum $\Delta e$ for the smaller values of $P_\mathrm{tr}$.
	
	{\it Unbound Region (Green)}---This region contains only a minimum value of $\Delta e$, as there is no maximum limit on $\Delta e$. As shown in Fig.~\ref{fig:theory_1} (right), we selected two EoSs from either side of the minimum $\Delta e$. The respective $M$--$R$ sequences show that TSs (indicated by the turquoise curve on the third branch) can form for much larger values of $\Delta e$. The plot shows the end of this region at a minimum value of $P_\mathrm{tr}$, where $M_\mathrm{TOV,2}=1.0~M_\odot$. However, that is only a restriction applied to the EoS survey in this study. The formation of TSs is still possible below the minimum value of $P_\mathrm{tr}$ accounted for in this study. That is, the locus of minimum $\Delta e$ will continue to exist below the limit of this region.

	In this article, we identify the loci of minimum and maximum $\Delta e$ with varying $P_\mathrm{tr}$ and fixed $e_\mathrm{tr}$ as the TS existence boundary. We note that the TS existence boundary can be computed for a given combination of the parameters---($e_\mathrm{tr}$, $c^2_\mathrm{s,QM}$). A point under the TS existence boundary represents a unique EoS (informing the values of $P_\mathrm{tr}$ and $\Delta e$ in addition to $e_\mathrm{tr}$ and $c^2_\mathrm{s,QM}$) that constructs TS solutions. Similar curves have been discussed before in the literature, but with a fixed hadronic EoS, which constrains $P_\mathrm{tr}$ given the choice of $e_\mathrm{tr}$ \cite{alford_generic_2013,christian_classifications_2018}. While treating $P_\mathrm{tr}$ and $e_\mathrm{tr}$ as free variables, we can explore a larger possibility of EoS, resulting in new features about the TS existence boundary that have not been discussed previously.

	\begin{table}
		\caption{Parameter space surveyed to determine the twin-star existence boundaries using the four-parameter Maxwell-CSS model. We systematically vary these parameters over the ranges and values indicated. The lower limit of $P_\mathrm{tr}$ is set to a value where such a central pressure would construct an NS of $1.0~M_\odot$. The maximum value of $P_\mathrm{tr}$ is determined by whether the respective density reaches the central density of $M_\mathrm{TOV}$ or $c_\mathrm{s}^2=1$, whichever occurs first, as indicated by (${}^*$) and (${}^{\#}$), respectively. For the PT region, the minimum $e_\mathrm{tr}$ surveyed in this study is $0.40\times10^{15}~\mathrm{g/cm}^3$ ($\approx 1.4\rho_\mathrm{sat}$). The CET sufficiently predicts that matter resides as hadronic matter up to $\sim 1.1\rho_\mathrm{sat}$~\cite{gandolfi_quantum_2009,tews_neutron_2013}. The maximum value of $e_\mathrm{tr}$ surveyed was limited to $1.50\times10^{15}~\mathrm{g/cm}^3$, as we found that EoSs with $e_\mathrm{tr}\gtrsim1.4\times10^{15}~\mathrm{g/cm}^3$ in our Maxwell-CSS model are incompatible with the observational constraints. The values of $c^2_\mathrm{s,QM}$ are reasonably surveyed between the conformal limit and causality limit, with a step size of $\sim 0.07$. We have found all the EoSs with $c^2_\mathrm{s,QM}\lesssim0.55$ in our Maxwell-CSS model to be incompatible with the observational constraints.}
		\centering
		
		\begin{tabular}{@{\hspace{0.0cm}}c@{\hspace{0.2cm}}c@{\hspace{0.2cm}}c@{\hspace{0.2cm}}c@{\hspace{0.0cm}}}
			
			\hline
			$e_\mathrm{tr}$ & $P_\mathrm{tr}$ & $\Delta e$ & $c^2_\mathrm{s,QM}$ \\
			$[\times10^{15}~\mathrm{g/cm}^3]$ & $[\times10^{35}~\mathrm{Dyne/cm}^2]$ & $[\times10^{15}~\mathrm{g/cm}^3]$ & $[c^2]$ \\
			\hline
			$0.40$ & $0.26 $--$ 0.98^{~\#}$ & $0.001$--$3.00$ & 0.33--1.0 \\
			$0.50$ & $0.36 $--$ 1.39^{~\#}$ & $0.001$--$3.00$ & 0.33--1.0 \\
			$0.60$ & $0.46 $--$ 1.82^{~*}$ & $0.001$--$3.00$ & 0.33--1.0 \\
			$0.90$ & $0.79 $--$ 3.21^{~*}$ & $0.001$--$3.00$ & 0.33--1.0 \\
			$1.20$ & $1.17 $--$ 4.70^{~*}$ & $0.001$--$3.00$ & 0.33--1.0 \\
			$1.50$ &  $1.60 $--$ 6.28^{~*}$ & $0.001$--$3.00$ & 0.33--1.0 \\
			\hline
		\end{tabular}
		\label{tab:parameter_ranges}
	\end{table}

	In Fig.~\ref{fig:theory_2} (left), we show that the TS existence boundaries heavily depend on the variation in ($e_\mathrm{tr}$, $c^2_\mathrm{s,QM}$). We find that an increase in $e_\mathrm{tr}$ leads to an increase in the size of the TS existence boundary, indicating a larger region is allowed for $P_\mathrm{tr}$ and $\Delta e$. For a fixed $e_\mathrm{tr}$, the lower values of $c^2_\mathrm{s,QM}$ form a smaller TS existence boundary with a lower value of $P^\mathrm{max}_\mathrm{tr}$. Interestingly, we observe a puncture on the tip of the TS existence boundary formed by ($e_\mathrm{tr,b}$, $c^2_\mathrm{s,QM1}$). Here, the loci of the minimum and maximum of $\Delta e$ do not meet, which is different from the expected picture discussed in Fig.~\ref{fig:theory_1}. We find that in such a case, the $P^\mathrm{max}_\mathrm{tr}$ would be beyond the causality limit, which is also the upper limit for the survey range of $P_\mathrm{tr}$. If we were to allow a survey of hadronic EoSs beyond the causality limit, the TS existence boundary would have been completed. As we lower the value of $e_\mathrm{tr}$, for instance ($e_\mathrm{tr,a}$, $c^2_\mathrm{s,QM1}$), we find that the TS existence boundary is even more incomplete, where we only find the locus of minimum $\Delta e$. Such scenarios result in $P_\mathrm{tr}$ at the causality limit being present in the unbound region (green) of Fig.~\ref{fig:theory_1}, and hence the incomplete TS existence boundary would not have a well-defined bound region (blue). Hence, for the incomplete TS existence boundaries, we define $P^\mathrm{max}_\mathrm{tr}$ at the causality limit. However, for smaller values of $c^2_\mathrm{s,QM}$, the TS existence boundary can be complete. For example, the TS existence boundary for ($e_\mathrm{tr,b}$, $c^2_\mathrm{s,QM2}$) is complete, whereas the TS existence boundary for ($e_\mathrm{tr,b}$, $c^2_\mathrm{s,QM1}$) is punctured. This effect is expected from the fact that smaller $c^2_\mathrm{s,QM}$ forms the TS existence boundary in the lower range of $P_\mathrm{tr}$. 
	
	In Fig.~\ref{fig:theory_2} (right), we compute the TS existence boundaries in $\Delta e/e_\mathrm{tr}-P_\mathrm{tr}/e_\mathrm{tr}$ axis plot, which has been widely used to study this subject in the past literature \cite{alford_generic_2013,christian_classifications_2018}. Since the Seidov lines align in this plot, unlike in Fig.~\ref{fig:theory_2} (left), the effect of $e_\mathrm{tr}$ and $c^2_\mathrm{s,QM}$ on TS existence boundaries can be compared directly. We demonstrate that the picture surrounding the TS existence boundaries discussed in this study agrees with the special cases presented in earlier studies. Therefore, the bifurcation of the region under the TS existence boundary based on categories of TSs, as defined in Ref.~\cite{christian_classifications_2018} (see Fig.~6 therein), remains consistent and applicable in our work. Since our ensemble of EoSs consists of Maxwell-type EoSs, it is important to note that the picture of TS existence boundaries will vary if hybrid EoSs with mixed phases are probed.

	\begin{figure*}
		\centering
		\includegraphics[scale=1.05]{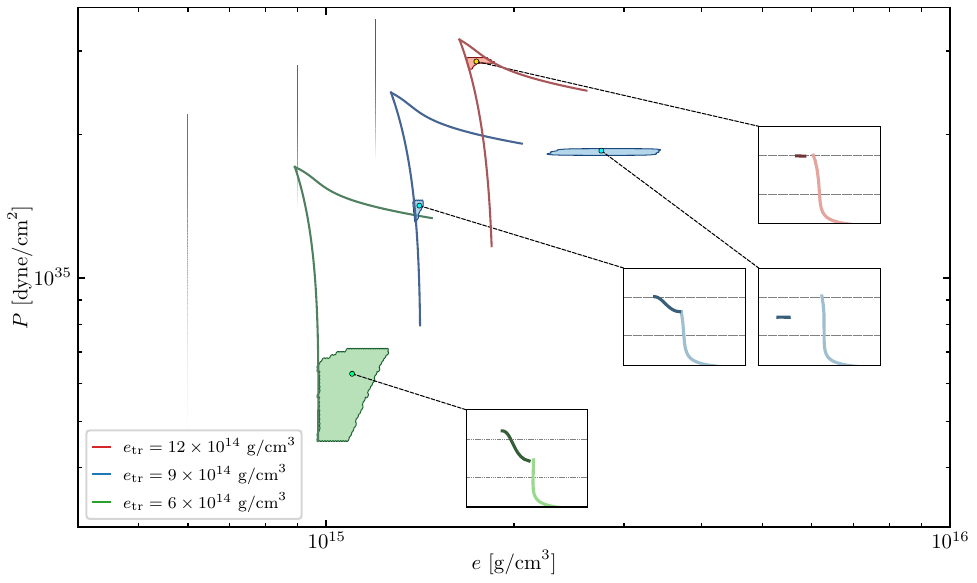}
		\caption{Allowed regions under the TS existence boundary for three different values of $e_\mathrm{tr}$, constrained from observational measurements ($2\sigma$  confidence). The vertical fading lines indicate the value of $e_\mathrm{tr}$. For all the EoSs, the quark part is described with $c^2_\mathrm{s,QM}=1.0$. The inset shows the $M$--$R$ sequences of the EoSs with ($P_\mathrm{tr}$, $\Delta e$) marked by circles inside the allowed region. Only the stable branches of $M$--$R$ sequences are shown, where the second (third) branch is indicated using light (dark) colour in the respective colour theme used for the $e_\mathrm{tr}$. Black dashed lines are marked at $1.0~M_\odot$ and $2.0~M_\odot$ for the identification of TS categories.}
		\label{fig:observation_1}
	\end{figure*}
	
	\begin{figure*}
		\includegraphics[scale = 1]{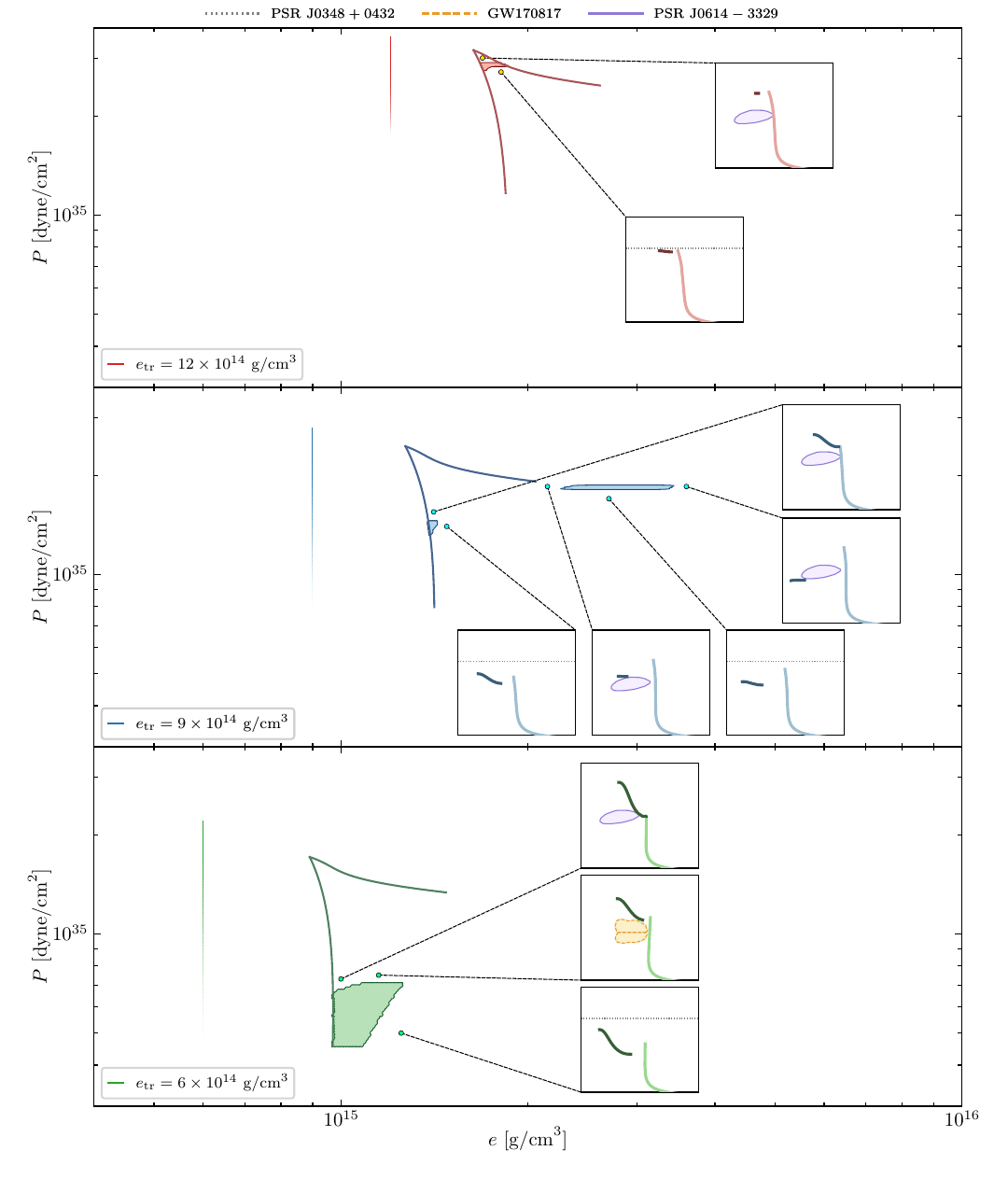}\vspace{-0.5cm}   
		\caption{Emphasis on particular observational measurements ($2\sigma$  confidence) that are responsible for constraining various regions under the TS existence boundary for three different values of $e_\mathrm{tr}$. The vertical fading lines indicate the value of $e_\mathrm{tr}$. The quark EoS description is set with $c^2_\mathrm{s,QM}=1.0$. The inset shows the $M$--$R$ sequences, along with the particular observational constraint that they violate. These $M$--$R$ sequences are constructed with the EoSs with ($P_\mathrm{tr}$, $\Delta e$) marked by circles outside the allowed region. Only the stable branches of $M$--$R$ sequences are shown, where the second (third) branch is indicated using light (dark) colour in the respective colour theme used for the $e_\mathrm{tr}$.}
		\label{fig:observation_2}
	\end{figure*}
	\begin{figure*}
		\centering
		\includegraphics[scale = 0.8]{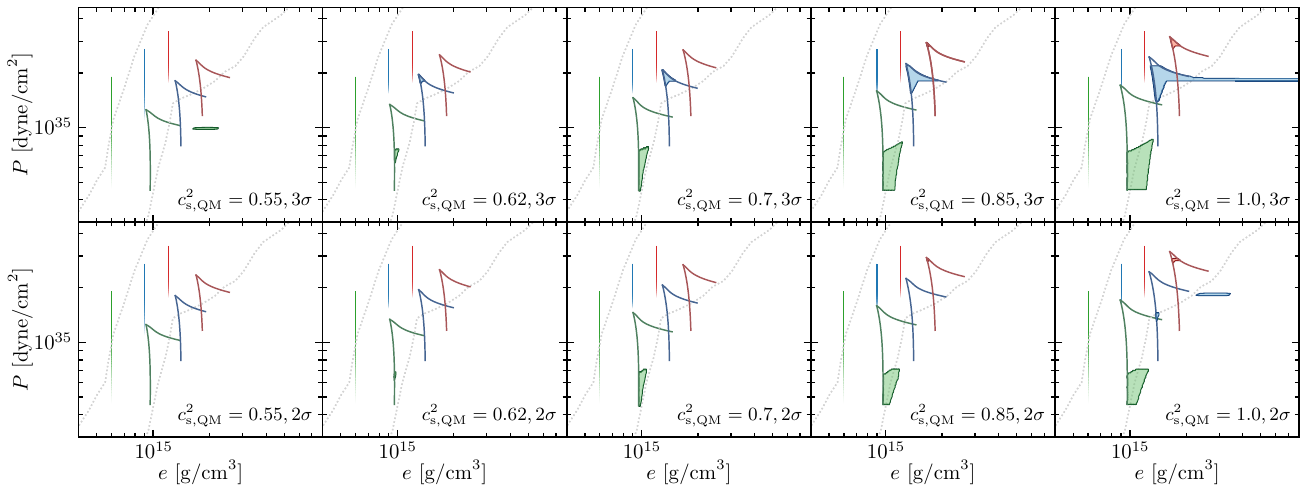}
		\caption{Change in the picture of allowed regions of $\Delta e$ with varying $c^2_\mathrm{s,QM}$ (0.55, 0.62, 0.7, 0.85 and 1.0) along with the imposition of the constraints from observational measurements at confidences $2\sigma$ and $3\sigma$. The choice of $c^2_\mathrm{s,QM}$ remains the same over a column, and the choice of confidence remains constant over a row. The background contour is taken from \cite{annala_evidence_2020}, indicating that the range of allowed EoSs in this study lies over (however, not completely within) the constrained region provided in the literature. The vertical fading lines indicate the value of $e_\mathrm{tr}$.}
		\label{fig:observation_3}
	\end{figure*}

	We observed that the TOV solutions can only put an upper limit on $P_\mathrm{tr}$ (if not by the causality limit). No upper limit is provided for $e_\mathrm{tr}$ or $\Delta e$ (at small values of $P_\mathrm{tr}$). Interestingly, we find that the current conglomerate of the observational measurements can provide further constraints on the picture formed within our four-parameter Maxwell-CSS model.
	
	\subsection{Constraints from recent observational measurements}
	\label{sec:obs}

	For developing further constraints from astrophysical point of view, we incorporated observational inference from NICER mass-radius measurements (PSR J0614--3329~\cite{mauviard_nicer_2025}, PSR J0437--4715~\cite{choudhury_nicer_2024}, PSR J0030+0451~\cite{riley_nicer_2019,miller_psr_2019,raaijmakers_nicer_2019}, and PSR J0740+6620~\cite{riley_nicer_2021,miller_radius_2021,raaijmakers_constraints_2021,fonseca_refined_2021}), the mass measurement of massive pulsar PSR J0348+0432~\cite{antoniadis_massive_2013}, and the tidal deformability from GW170817~\cite{abbott_gw170817_2018}. Additional constraints have been implemented at lower densities from the CET \cite{altiparmak_sound_2022}. A table of our survey parameters is noted in Table~\ref{tab:parameter_ranges}.

	These constraints are imposed by extracting the contours at different confidence intervals, followed by a hard cut filtering of the ensemble of EoSs. Since the primary objective of this work is not to identify the most probable regions of the parameter space through a Bayesian framework, but rather to perform a systematic exploration of our Maxwell-CSS model four-parameter space and find the regions that remain compatible with the imposed observational criteria, our study establishes the boundaries on the parameter space within the chosen model. Because observational likelihoods are converted into binary acceptance regions, that is, points just inside and far inside a confidence contour receive the same status, the degree of agreement between an EoS and the observations is not quantified in our analysis.

	Fig.~\ref{fig:observation_1} shows that the allowed region under a TS existence boundary can be heavily constrained using these observational constraints (using $2\sigma$ confidence), and only selective regions of ($P_\mathrm{tr}$, $\Delta e$) are allowed for each value of $e_\mathrm{tr}$. We present this picture using three specific values of $e_\mathrm{tr}$ (with $c^2_\mathrm{s,QM}=1.0$), which sufficiently explain the constraints received from observational measurements at different regimes of transition densities:
	\begin{itemize}
		\item {\it small}: $e_\mathrm{tr}=6\times 10^{14}~\mathrm{g/cm}^3$, $\rho_\mathrm{tr}\approx 2\rho_\mathrm{sat}$
		\item {\it intermediate}: $e_\mathrm{tr}=9\times 10^{14}~\mathrm{g/cm}^3$, $\rho_\mathrm{tr}\approx 3\rho_\mathrm{sat}$
		\item {\it large}: $e_\mathrm{tr}=12\times 10^{14}~\mathrm{g/cm}^3$, $\rho_\mathrm{tr}\approx 4\rho_\mathrm{sat}$
	\end{itemize}

	In Fig.~\ref{fig:observation_2}, we emphasise that different observational measurements become relevant to constrain different regions under the TS existence boundaries, and that different combinations of observational measurements come into play at different regimes of $e_\mathrm{tr}$. Firstly, we show that, for all three values of $e_\mathrm{tr}$, the maximum values of $\Delta e$ are constrained by the current observational inferences (see more discussion in Sec.~\ref{sec:limits}). In particular, GW170817, PSR J0348+0432 and PSR J0614--3329 assist in constraining the respective maximum $\Delta e$ at small, intermediate and large values of  $e_\mathrm{tr}$. Since larger values of $\Delta e$ result in more compact third branches (smaller radii), the maximum $\Delta e$ is bound to be constrained by the observational measurements at some point, as all of the contours have a minimum value in the uncertainty of the radii. It is subject to the fact that we did not find an EoS using our formalism that forms a second branch satisfying all the constraints by itself and then forms a third branch in the regime of the unbound region (green) discussed in Fig.~\ref{fig:theory_1} (however, such a condition can be found using $3\sigma$ confidence, see discussion in Sec.~\ref{sec:limits}). In such a case, finding a maximum limit on $\Delta e$ using observational constraints would not be possible. The maximum values of $\Delta e$ ($\Delta \rho$) were noted to be $6.52\times 10^{14}~\mathrm{g/cm}^3$ ($1.98\rho_\mathrm{sat}$), $24.95\times 10^{14}~\mathrm{g/cm}^3$ ($6.59\rho_\mathrm{sat}$) and $6.53\times 10^{14}~\mathrm{g/cm}^3$ ($1.63\rho_\mathrm{sat}$) at small, intermediate and large $e_\mathrm{tr}$, respectively.

	Secondly, in Fig.~\ref{fig:observation_1}, we observe that, even though the larger $e_\mathrm{tr}$ results in larger TS existence boundaries, observational measurements tend to constrain these regions more tightly. We find that the collection of allowed (satisfying causality) hadronic EoSs at larger $e_\mathrm{tr}$ is softer, resulting in lower $M_\mathrm{TOV}$. As a result, at extremely large values of $e_\mathrm{tr}$, the hybrid EoSs consisting of such hadronic parts won't be stiff enough to satisfy the mass constraint from PSR J0348+0432. Hence, there exists a maximum value of $e_\mathrm{tr}$ for the TS solution in light of current observational constraints (see more discussion in Sec.~\ref{sec:limits}). 
	
	For small values of $e_\mathrm{tr}$, the region under the TS existence boundary is constrained to lower values of $P_\mathrm{tr}$, as observed in Fig.~\ref{fig:observation_1}. The selected EoS in this region indicates that the allowed range of $M$--$R$ sequences can only span categories III and IV. In Fig.~\ref{fig:observation_2}, we show that the mass measurement of PSR J0348+0432 constrains $\Delta e$, and GW170817 and PSR J0614--3329 are responsible for constraining $P_\mathrm{tr}$. Importantly, the latter observations rule out the region of EoSs right below the tip of the TS existence boundary, which mainly constructs category I TSs. The involved constraints in this regime demand a softer hadronic part in the EoSs such that the second branch follows the constraints, as opposed to the requirement of stiffer hadronic EoSs to construct massive TSs of category I.

	For intermediate values of $e_\mathrm{tr}$, we observe two disconnected regions under the TS existence boundary. These are divided into lower ($P_\mathrm{tr}$, $\Delta e$) and higher ($P_\mathrm{tr}$, $\Delta e$) regions, which are strongly associated with the construction of TSs of category III and II, respectively. In Fig.~\ref{fig:observation_2}, we show that the mass-radius regions inferred from PSR J0614--3329 and PSR J0348+0432 are mainly responsible for ruling out these EoSs. The lower ($P_\mathrm{tr}$, $\Delta e$) region is drastically smaller in comparison to the allowed region for small $e_\mathrm{tr}$. It indicates that category IV TSs are ruled out at intermediate $e_\mathrm{tr}$ values. Here, the $\Delta e$ and lower $P_\mathrm{tr}$ constraint come from the violation of the mass constraint from PSR J0348+0432, whereas the higher $P_\mathrm{tr}$ constraint comes from the violation of PSR J0614--3329. On the other hand, for higher ($P_\mathrm{tr}$, $\Delta e$), $\Delta e$ is constrained on either side due to the violation of PSR J0614--3329. Similarly, category I TSs continue to be ruled out at intermediate $e_\mathrm{tr}$.

	\begin{figure*}
		\centering
		\includegraphics[scale=1.05]{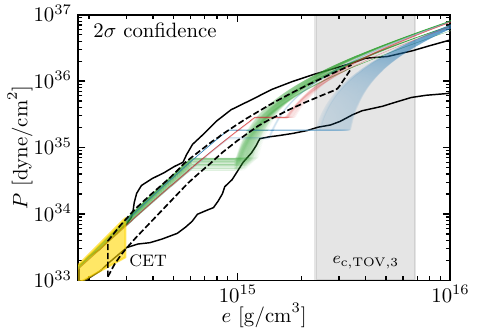}
		\includegraphics[scale=1.05]{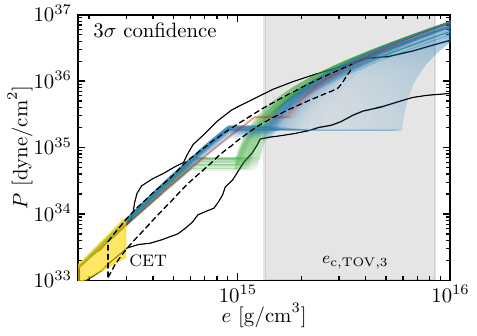}
		\caption{EoSs curves, which are presented in Fig.~\ref{fig:observation_3} for confidence intervals $2\sigma$ (on the left panel) and $3\sigma$ (on the right panel) against the EoS contour (black solid lines) given by \cite{annala_evidence_2020}. An additional contour is added indicating the posterior domain of the nucleonic EoS DDB from Fig.~1 of ref.~\cite{sanson_hyperonic_2026} to compare our surviving parameter space for hadronic EoS with one of the known hadronic EoS from microphysics. The lower density is constrained by the CET band (yellow region) given by \cite{altiparmak_sound_2022}. The grey region indicates the range of central energy densities ($e_\mathrm{c}$) of maximum mass on the third branch ($M_\mathrm{TOV,3}$) indicating that all EoSs can be smoothly matched to pQCD after this value has been reached.}
		\label{fig:observation_4}
	\end{figure*}

	Interestingly, for large values of $e_\mathrm{tr}$, only the region near the tip of the TS existence boundary is allowed, as the rest of the region continues to violate constraints from PSR J0614--3329 and PSR J0348+0432. Since the set of allowed hadronic EoSs is rather soft, only a few hybrid EoSs can produce massive NSs ($\gtrsim 2M_\odot$) while satisfying all the constraints. Therefore, TSs of category I are only possible for larger $e_\mathrm{tr}$ values, while the rest of the categories of TSs are strongly ruled out.
	
	Since the TS existence boundaries are also dictated by $c^2_\mathrm{s,QM}$ (as discussed in Sec.~\ref{sec:theory}), the allowed regions from observations will also be affected similarly. In Fig.~\ref{fig:observation_3}, we summarise the effect of $c^2_\mathrm{s,QM}$ in the picture. Since the resulting TS existence boundaries are smaller with a decrease in $c^2_\mathrm{s,QM}$, the observationally constrained regions are also tighter (from the right to the left panel). 
	
	Additionally, we inform the further inclusion of EoSs when $3\sigma$ confidence of observational inferences is considered. As expected, an increase in the confidence interval (from the lower to the upper panel) widens the set of allowed EoSs. Interestingly, if $3\sigma$ confidence is imposed to constrain EoSs with intermediate values of $e_\mathrm{tr}$, we observe that the EoSs with category I TSs are included at values of $c^2_\mathrm{s,QM} \gtrsim 0.6$, which remained excluded in our examination while considering $2\sigma$ confidence of observational data. For large $e_\mathrm{tr}$, the inferences are unaffected when $3\sigma$ confidence is imposed; that is, category I TSs vanish when EoSs with $c^2_\mathrm{s,QM} \lesssim 0.7$ are considered. Secondly, at intermediate $e_\mathrm{tr}$ with $c^2_\mathrm{s,QM}=1.0$, we observe that a slim range of $P_\mathrm{tr}$ is allowed with $\Delta e$ being unconstrained. These hybrid EoSs construct a second branch that satisfies all the constraints while having its $P_\mathrm{tr}$ in the unbound region (green), as discussed in Fig.~\ref{fig:theory_1}. However, in the interest of extending these hybrid EoSs to match the known behaviour of matter at pQCD regime (see Refs.~\cite{annala_evidence_2020,altiparmak_sound_2022,blomqvist_strong_2026} for such consistency treatments to agnostic EoSs), an ultimate upper bound can be drawn on $\Delta e$ ($\Delta\rho$) to be $\sim5\times10^{15}~\mathrm{g/cm}^3$ ($\sim  13\rho_\mathrm{sat}$). Thirdly, we observe that EoSs need very stiff quark EoSs ($c^2_\mathrm{s,QM}\gtrsim0.85$) to construct TSs of category II, which remains true when both confidence intervals are considered. As an exception, category II TSs reappear for $c^2_\mathrm{s,QM} \approx 0.55$, when the second branch satisfies constraints from PSR 437--4715 using $3\sigma$ confidence. Lastly, we found that category IV TSs can only be constructed using small $e_\mathrm{tr}$ combined with the quark part of $c^2_\mathrm{s,QM} \gtrsim 0.62$, which remains consistent over both confidence intervals.

	\begin{table}[tbh!]
		\caption{Maximum transition energy density ($e^\mathrm{max}_\mathrm{tr}$) and respective maximum transition rest-mass density ($\rho^\mathrm{max}_\mathrm{tr}$) for various stiffnesses of quark matter EoSs using $c^2_\mathrm{s,QM}$. The values of $\Delta e$ and $\Delta \rho$ are also noted for these selected EoSs, along with their category of TSs. This computation is performed considering $2\sigma$ confidence level.}
		\centering
		\begin{tabular}{@{\hspace{0.0cm}}c@{\hspace{0.3cm}}c@{\hspace{0.3cm}}c@{\hspace{0.3cm}}c@{\hspace{0.3cm}}c@{\hspace{0.3cm}}c@{\hspace{0.0cm}}}
			\hline
			$c^2_\mathrm{s,QM}$ & $e^\mathrm{max}_\mathrm{tr}$ & $\Delta e$ & $\rho^\mathrm{max}_\mathrm{tr}$ & $\Delta\rho$ & Cat.\\
			$[c^2]$ & $[\times10^{15}~\mathrm{g/cm}^3]$ & $[\times10^{15}~\mathrm{g/cm}^3]$ & $ [\rho_\mathrm{sat}]$ & $ [\rho_\mathrm{sat}]$ &\\
			
			\hline
			$1.0$  & $1.40$ & $0.48$ & $4.03$ & $1.14$ & I   \\
			$0.85$ & $1.27$ & $0.46$ & $3.70$ & $1.14$ & I   \\ 
			$0.55$ & $0.54$ & $0.31$ & $1.87$ & $1.15$ & III \\ 
			\hline
		\end{tabular}
		\label{tab:e_tr_max_cs2}
	\end{table}

	\begin{table}[tbh!]
		\caption{Maximum transition energy density ($e^\mathrm{max}_\mathrm{tr}$) and respective maximum transition rest-mass density ($\rho^\mathrm{max}_\mathrm{tr}$) under different confidence levels of constraints from observational inferences. The values of $\Delta e$ and $\Delta \rho$ are also noted for these selected EoSs, along with their category of TSs. This computation is performed considering the stiffest quark EoS ($c^2_\mathrm{s,QM}=1.0$).}
		\centering
		\begin{tabular}{@{\hspace{0.0cm}}c@{\hspace{0.2cm}}c@{\hspace{0.2cm}}c@{\hspace{0.2cm}}c@{\hspace{0.2cm}}c@{\hspace{0.2cm}}c@{\hspace{0.0cm}}}
			\hline
			Conf. & $e^\mathrm{max}_\mathrm{tr}$ & $\Delta e$ & $\rho^\mathrm{max}_\mathrm{tr}$ & $\Delta \rho$ & Cat.\\
			& $[\times10^{15}~\mathrm{g/cm}^3]$ & $[\times10^{15}~\mathrm{g/cm}^3]$ & $[\rho_\mathrm{sat}]$ & $ [\rho_\mathrm{sat}]$ &\\
			\hline
			$3\sigma$ & $1.40$ & $0.48$ & $4.03$ & $1.14$ & I\\ 
			$2\sigma$ & $1.40$ & $0.48$ & $4.03$ & $1.14$ & I\\           
			
			\hline
		\end{tabular}
		\label{tab:e_tr_max}
	\end{table}
	
	\begin{table}[tbh!]
		\caption{Maximum strength of PT discontinuity in energy density ($\Delta e_\mathrm{max}$) and respective rest-mass density ($\Delta \rho_\mathrm{max}$) for various stiffnesses of quark matter EoSs using $c^2_\mathrm{s,QM}$. The values of $e_\mathrm{tr}$ and $\rho_\mathrm{tr}$ are also noted for the allowed EoS along with their category of TSs. This computation is performed considering $2\sigma$ confidence level.}
		\centering
		\begin{tabular}{@{\hspace{0.0cm}}c@{\hspace{0.2cm}}c@{\hspace{0.2cm}}c@{\hspace{0.2cm}}c@{\hspace{0.2cm}}c@{\hspace{0.2cm}}c@{\hspace{0.2cm}}c@{\hspace{0.0cm}}}
			\hline
			$c^2_\mathrm{s,QM}$ & $\Delta e_{\mathrm{max}}$ & $\Delta \rho_{\mathrm{max}}$ &$e_\mathrm{tr}$  & $\rho_\mathrm{tr}$ & Cat.\\
			$[c^2]$ & $[\times10^{15}~\mathrm{g/cm}^3]$ & $[\rho_\mathrm{sat}]$ & $ [\times10^{15}~\mathrm{g/cm}^3]$ & $ [\rho_\mathrm{sat}]$ &\\
			\hline
			$1.0$  & $2.94$  & $7.76$  & $0.90$ & $2.92$  & II  \\ 
			$0.85$ & $2.35$  & $6.29$  & $0.88$ & $2.85$  & II  \\ 
			$0.55$ & $0.34$  & $1.26$  & $0.54$ & $2.00$  & III \\ 
			\hline
		\end{tabular}
		\label{tab:del_e_max_cs2}
	\end{table}

	\begin{table}[tbh!]
		\caption{Maximum strength of PT discontinuity in energy density ($\Delta e_\mathrm{max}$) and respective rest-mass density ($\Delta \rho_\mathrm{max}$) under different confidence levels of constraints from observational inferences. The values of $e_\mathrm{tr}$ and $\rho_\mathrm{tr}$ are also noted for the allowed EoS along with their category of TSs. This computation is done for the stiffest quark EoS ($c^2_\mathrm{s,QM}=1.0$).}
		\centering
		\begin{tabular}{@{\hspace{0.0cm}}c@{\hspace{0.2cm}}c@{\hspace{0.2cm}}c@{\hspace{0.2cm}}c@{\hspace{0.2cm}}c@{\hspace{0.2cm}}c@{\hspace{0.2cm}}c@{\hspace{0.0cm}}}
			\hline
			Conf. & $\Delta e_{\mathrm{max}}$ & $\Delta \rho_{\mathrm{max}}$ &$e_\mathrm{tr}$  & $\rho_\mathrm{tr}$ & Cat.\\
			& $[\times10^{15}~\mathrm{g/cm}^3]$ & $[\rho_\mathrm{sat}]$ & $ [\times10^{15}~\mathrm{g/cm}^3]$ & $ [\rho_\mathrm{sat}]$ &\\
			\hline
			$3\sigma$ & $4.97$ & $13.13$ & $0.90$ & $2.92$ & II\\
			$2\sigma$ & $2.94$ & $7.76$  & $0.90$ & $2.92$ & II\\      
			\hline
		\end{tabular}
		\label{tab:del_e}
	\end{table}

	\begin{table} [tbh!]
		\caption{Maximum mass of TSs ($M^\mathrm{max}_\mathrm{TS}$) and their categories, for various stiffnesses of quark matter EoSs using $c^2_\mathrm{s,QM}$. The values of $\Delta e$ and $\Delta \rho$ are also noted for the allowed EoS that construct the stellar configurations supporting $M^\mathrm{max}_\mathrm{TS}$. This computation is performed considering $2\sigma$ confidence level.}
		\centering
		\begin{tabular}{@{\hspace{0.0cm}}c@{\hspace{0.2cm}}c@{\hspace{0.2cm}}c@{\hspace{0.2cm}}c@{\hspace{0.2cm}}c@{\hspace{0.2cm}}c@{\hspace{0.2cm}}c@{\hspace{0.0cm}}}
			\hline
			$c^2_\mathrm{s,QM}$ & $M^\mathrm{max}_\mathrm{TS}$ & $e_\mathrm{tr}$ & $\Delta e$ & $\rho_\mathrm{tr}$ & $\Delta\rho$ & Cat.\\
			$[c^2]$& $[M_\odot]$& $[\times10^{15}\mathrm{g/cm}^3]$ & $[\times10^{15}\mathrm{g/cm}^3]$ & $ [\rho_\mathrm{sat}]$ & $ [\rho_\mathrm{sat}]$ &\\
			\hline
			$1.0$  &  $2.05$ & $1.26$  & $0.45$ & $3.66$ & $1.10$ & I \\
			$0.85$ &  $2.02$ & $1.21$  & $0.45$ & $3.55$ & $1.10$ & I \\
			$0.55$ &  $1.32$ & $0.54$  & $0.35$ & $2.00$ & $1.29$ & III\\
			\hline
		\end{tabular}
		\label{tab:m_max_cs2}
	\end{table}

	\begin{table}[tbh!]
		\caption{Maximum mass of TSs ($M^\mathrm{max}_\mathrm{TS}$) and their categories, under different confidence levels of constraints from observational inferences. The values of $\Delta e$ and $\Delta \rho$ are also noted for the allowed EoS that construct the stellar configurations supporting $M^\mathrm{max}_\mathrm{TS}$. This computation is performed considering the stiffest quark EoS ($c^2_\mathrm{s,QM}=1.0$).}
		\centering
		\begin{tabular}{@{\hspace{0.0cm}}c@{\hspace{0.2cm}}c@{\hspace{0.2cm}}c@{\hspace{0.2cm}}c@{\hspace{0.2cm}}c@{\hspace{0.2cm}}c@{\hspace{0.2cm}}c@{\hspace{0.0cm}}}
			\hline
			Conf. & $M^\mathrm{max}_\mathrm{TS}$ & $e_\mathrm{tr}$ & $\Delta e$ & $\rho_\mathrm{tr}$ & $\Delta \rho$ & Cat.\\
			& $[M_\odot]$& $[\times10^{15}\mathrm{g/cm}^3]$ & $[\times10^{15}\mathrm{g/cm}^3]$ & $ [\rho_\mathrm{sat}]$ & $ [\rho_\mathrm{sat}]$ &\\
			\hline
			$3\sigma$ & $2.32$ & $0.97$ & $0.39$ & $2.92$ & $0.94$ & I\\
			$2\sigma$ & $2.05$ & $1.26$ & $0.45$ & $3.66$ & $1.01$ & I\\ 
			\hline
		\end{tabular}
		\label{tab:m_max}
	\end{table}

	We did not find any EoSs that satisfy the observational constraints at the ($1\sigma$) confidence level within the EoS formalism considered in this work. However, this conclusion is limited to our specific parametrisation, and further investigations employing alternative EoS formalisms are needed to determine whether it holds more generally.
	
	Figure~\ref{fig:observation_4} shows the constrained EoS trajectories for three selected values of $e_\mathrm{tr}$, which indicates that EoSs constructing TSs of category II fall outside the contour from previous studies. It is important to note that previous studies that extracted the constrained EoS contours constructed ensembles of EoSs using the speed of sound interpolation method \cite{annala_evidence_2020,altiparmak_sound_2022}, a technique that is unlikely to produce EoSs with strong PT unless specified. It is evident from our study that EoSs with strong PT can exist under current constraints, and importantly, they form TSs of category II, which can serve as smoking gun signatures of strong PT. An additional contour is added indicating the posterior domain of the nucleonic EoS DDB from Fig.~1 of ref.~\cite{sanson_hyperonic_2026}, to compare our surviving parameter space for hadronic EoS with one of the known hadronic EoS from microphysics. Although the posterior added here is for a selected EoS, Fig.~1 of ref.~\cite{sanson_hyperonic_2026} suggests that other nucleonic EoS posteriors also follow similar trajectories and hence the surviving parameter space for hadronic EoS stands consistent with other hadronic EoS models.

	\subsection{Limits on properties of PT and maximum mass of TSs}
	\label{sec:limits}

	In Table~\ref{tab:e_tr_max_cs2}, we note the EoSs with the maximum value of $e_\mathrm{tr}$ ($e_\mathrm{tr}^\mathrm{max}$), for varying stiffness of quark EoS. It is intuitive to understand that the EoSs that form category I TSs at large $e_\mathrm{tr}$ ($\approx 4\rho_\mathrm{sat}$) are responsible for setting the upper bound on $e_\mathrm{tr}$. Such hybrid EoSs are soft enough to construct second branches that follow all the constraints ($2\sigma$ confidence level) while supporting massive TSs on the third branch. Since $P_\mathrm{tr}$ of these EoSs lie in the bound region (blue) of the TS existence boundary, the maximum of $\Delta e$ is strictly determined from TOV solutions. In the case of softer quark EoSs, the hybrid EoSs do not support massive TSs to satisfy the mass constraint from PSR J0348+0432. It can be anticipated from Fig.~\ref{fig:theory_2} where we show that smaller $c^2_\mathrm{s,QM}$ sets smaller $P_\mathrm{tr}^\mathrm{max}$. It suggests that a second branch supports massive NSs ($\gtrsim 2.0~M_\odot$) that will have central pressures higher than $P_\mathrm{tr}^\mathrm{max}$, and for such EoSs, TS solutions would not exist. In such cases, $e_\mathrm{tr}^\mathrm{max}$ is achieved via EoSs with TSs of category III, which is the reason behind the $e_\mathrm{tr}^\mathrm{max}$ being drastically smaller than we found for stiffer quark EoSs. In Table~\ref{tab:e_tr_max}, we find the $e_\mathrm{tr}^\mathrm{max}$ for various confidence intervals using the stiffest quark EoS. The results remain unchanged for $2\sigma$ and $3\sigma$ confidence levels, as these results are mainly bound by the mass constraint from PSR J0348+0432.

	Similarly, in Tables \ref{tab:del_e_max_cs2} and \ref{tab:del_e}, we note the EoSs with the maximum value of $\Delta e$ ($\Delta e_\mathrm{max}$). These inferences are made from EoSs that construct TSs of category II (III) for stiff (soft) quark EoS. For a stiffer quark EoS, the hybrid EoSs construct a second branch that satisfies all observations except for PSR J0614--3329. The third branch of category II kind then satisfies PSR J0614--3329, resulting in the inclusion of such an EoS with large $\Delta e$. For a softer quark EoS, the hybrid EoSs construct a second branch that satisfies the observation PSR J0740+6630, whereas the third branch of category II kind then satisfies the rest of the constraints. 
	
	In Table~\ref{tab:m_max_cs2}, we note the maximum mass of the TSs ($M^\mathrm{max}_\mathrm{TS}$) considering various stiffnesses of quark matter EoS (not to be confused with $M_\mathrm{TOV,3}$ as defined in Sec.~\ref{sec:eos}). For soft quark EoSs, we observe that the $M^\mathrm{max}_\mathrm{TS}$ is achieved for $\rho_\mathrm{tr}\sim2\rho_\mathrm{sat}$, which is close to the limit of our analysis. Higher $M^\mathrm{max}_\mathrm{TS}$ can be found for lower $\rho_\mathrm{tr}$, but it will be very low for a hadron-to-quark PT. In Table~\ref{tab:m_max}, we note the $M^\mathrm{max}_\mathrm{TS}$ considering various confidence intervals of observational constraints. For $2\sigma$ and $3\sigma$ confidence, we observe that $M^\mathrm{max}_\mathrm{TS}>2.0~M_\odot$, constructed out of category I TSs.  Interestingly, for $3\sigma$ confidence, an $M^\mathrm{max}_\mathrm{TS}>2.3~M_\odot$ is possible, which is substantially heavier than previously discussed. 
	
	\section{Conclusions}
	\label{sec:Summ}

	We systematically investigated TS solutions using a four-parameter Maxwell-CSS construction, characterised by $e_\mathrm{tr}$, $P_\mathrm{tr}$, $\Delta e$, and $c^2_{\mathrm{s,QM}}$. Within the explored parameter space, causal TOV solutions divide $P_\mathrm{tr}$, at fixed $(e_\mathrm{tr},c^2_{\mathrm{s,QM}})$, into solution-less, bound, and unbound regions. The loci of the minimum and maximum $\Delta e$ define the TS existence boundary, whose enclosed points in the $(P_\mathrm{tr},\Delta e)$ plane correspond to hybrid EoSs supporting TSs. Increasing either $e_\mathrm{tr}$ or $c^2_{\mathrm{s,QM}}$ enlarges this region, although causality can truncate the boundary for certain parameter combinations.
	
	For complete boundaries, the tip identifies the unique $(P_\mathrm{tr},\Delta e)$ at which the twin branch first appears for a given hadronic--quark EoS combination. At lower $P_\mathrm{tr}$, multiple values of $\Delta e$ can produce distinct $M$--$R$ sequences with twin branches. This structure is not universal: for some $e_\mathrm{tr}$, several physically allowed twin branches with different $\Delta e$ emerge simultaneously at the same $P_\mathrm{tr}$, and no unique onset discontinuity exists. The corresponding boundary has no tip because the hadronic EoS reaches the causality limit before the required high-pressure region becomes accessible. This previously unidentified feature emerges from our systematic exploration of the full four-dimensional parameter space.
	
	Causal TOV solutions alone constrain $P_\mathrm{tr}$ from above, unless causality imposes the limit, but provide no upper bound on $e_\mathrm{tr}$ or on $\Delta e$ at low $P_\mathrm{tr}$. We therefore applied current observational constraints at the $2\sigma$ level to representative small, intermediate, and large values of $e_\mathrm{tr}$, fixing $c^2_{\mathrm{s,QM}}=1$. The allowed region becomes progressively narrower with increasing $e_\mathrm{tr}$, and $\Delta e$ acquires an upper bound throughout. An upper limit on $e_\mathrm{tr}$ also arises because large transition densities allow softer hadronic EoSs that cannot support massive NSs with $M\gtrsim 2 \,M_\odot$. The observations additionally impose a maximum TS mass.
	
	For the maximally stiff quark EoS, we report conditional upper bounds on $e_\mathrm{tr}$, $\Delta e$, and $M_\mathrm{TS}$ from the extrema surviving the adopted $2\sigma$ hard cut. These bounds are conditional on both the assumed quark EoS and the hard-cut procedure. At $3\sigma$, no upper bound on $\Delta e$ is obtained for large $e_\mathrm{tr}$; within these relaxed constraints, a narrow range of $P_\mathrm{tr}$ in the unbound region can still produce a second $M$--$R$ branch consistent with all observations.
	
	Under the $2\sigma$ hard cut, category-I TSs require large $e_\mathrm{tr}$ and $P_\mathrm{tr}$, together with a stiff quark EoS. Category II is restricted to intermediate $e_\mathrm{tr}$ and a very stiff quark EoS. Category III occurs either for intermediate $e_\mathrm{tr}$ with a maximally stiff quark EoS or for small $e_\mathrm{tr}$ with an intermediate-to-stiff quark EoS. Category IV is confined to small $e_\mathrm{tr}$ and $P_\mathrm{tr}$, with a significantly stiff quark EoS.
	
	Observations of TSs, particularly massive configurations near ($M^\mathrm{max}_\mathrm{TS}$), could provide smoking-gun signatures of a hadron–quark PT in NS cores and place strong constraints on ($e\mathrm{tr}$) and ($\Delta e$). Future studies should incorporate rotation and magnetic fields and consider PT constructions with a non-vanishing speed of sound, including finite-size effects associated with varying hadron–quark surface tension \cite{maslov_hybrid_2019,blaschke_mixing_2020,abgaryan_two_2018,yasutake_finite-size_2014,thi_uncertainties_2021,ju_hadron-quark_2021}. A complementary analysis using speed of sound parametrisations for both hadronic and quark EoSs, rather than the piecewise-polytropic CSS framework adopted here, has recently been performed in Ref.~\cite{blomqvist_strong_2026}. Moreover, slow hadron–quark conversion can dynamically stabilise TSs on the conventionally unstable branch \cite{ranea-sandoval_constraining_2022, lugones_model-agnostic_2023, ranea-sandoval_asteroseismology_2023, mariani_could_2024, mariani_how_2025, mariani_systematic_2026}, potentially enlarging the parameter space compatible with current observations.
	
	Recent likelihood-based analyses, such as \cite{Brandes_2025, montefusco_asymptotically_2026}, incorporate the latest NICER measurements through likelihood or posterior weighting, thereby retaining information on the relative degree of agreement between different EoS models and the observations. In contrast, our analysis implements the observational constraints as binary compatibility criteria. This approach does not assign posterior weights or quantify the relative likelihood of the compatible EoS. Its purpose is instead to provide a transparent identification of the existence and compatibility boundaries of the considered EoS parameter space. The Bayesian quantification of observability of TSs and the subsequent detectability of PT in NSs has been discussed in recent studies of \cite{grundler_bayesian_2025, mondal_detectability_2023}.
	
	We finally emphasise the particular significance of category-II TSs, which constitute only a small fraction of viable TS EoSs and are disfavoured by the Bayesian analysis of Ref.~\cite{blomqvist_strong_2026}. However, such inferences rely exclusively on hydrostatic-equilibrium solutions and neglect the dynamical response of these solutions to astrophysical perturbations. A dedicated study \cite{haque_dynamical_2026} shows that, under small perturbations, category-II TSs preferentially migrate to hadronic configurations on the neighbouring branch or collapse into black holes.\\

	\section*{Acknowledgements}
	
	It is a pleasure to thank Sagnik Chatterjee and Luciano Rezzolla for their crucial insights into this work. The authors also express their gratitude to the Deucalion HPC in Portugal for its support in the Advanced Computing Project 2025.00067.CPCA A3, RNCA (Rede Nacional de Computação Avançada), financed by the FCT (Fundação para a Ciência e a Tecnologia, IP, Portugal). The test setups were completed on the National Supercomputing Mission (NSM) resource ‘PARAM Ganga’ at IIT Roorkee, which is implemented by C-DAC and supported by the Ministry of Electronics and Information Technology (MeitY) and the Department of Science and Technology (DST), Government of India. The small-scale tests were conducted at the Bhaskara HPC in IISER Bhopal. SH and RM acknowledge the SERB grant: CRG/2022/000663. S.H. is supported by Research Associateship under the Anusandhan National Research Foundation, India, ARG Grant (No.ANRF/ARG/2025/005821/PS). SH acknowledges the International Centre for Theoretical Sciences (ICTS), Bengaluru, funded by the Department of Atomic Energy, Government of India, for hosting and supporting his visit (under project no: RTI4019) and participation in the program - Summer School on Gravitational-Wave Astronomy 2026 (code: ICTS/SSGWA2026/07).
	
	This research work is done using the packages---{\sc Numpy}~\cite{harris_array_2020}, {\sc SciPy}~\cite{virtanen_scipy_2020}, {\sc Matplotlib}~\cite{hunter_matplotlib_2007}, {\sc Seaborn}~\cite{waskom_seaborn_2021}, {\sc Jupyter}~\cite{kluyver_jupyter_2016}, {\sc Numba}~\cite{lam_numba_2015} and {\sc OpenCV}~\cite{bradski_opencv_2000}.

	\bibliography{PRD}
	
\end{document}